\documentclass[twocolumn,showpacs,prl,superscriptaddress]{revtex4}

\usepackage{graphicx}
\usepackage[usenames]{color}
\usepackage{amssymb}
\usepackage{amsmath}
\usepackage{amsfonts}
\usepackage{mathrsfs}

\renewcommand{\epsilon}{\varepsilon}
\newcommand{\figurewidth}{0.39\textwidth}

\begin{document}
\title{Influence of polymer-pore interactions on translocation}
\author{Kaifu Luo }
\altaffiliation[]{
Author to whom the correspondence should be addressed}
\email{luokaifu@yahoo.com}
\affiliation{Laboratory of Physics, Helsinki University of Technology,
P.O. Box 1100, FIN-02015 TKK, Espoo, Finland}
\author{Tapio Ala-Nissila}
\affiliation{Laboratory of Physics, Helsinki University of Technology,
P.O. Box 1100, FIN-02015 TKK, Espoo, Finland}
\affiliation{Department of Physics, Box 1843, Brown University, Providence,
Rhode Island 02912-1843, USA}
\author{See-Chen Ying}
\affiliation{Department of Physics, Box 1843, Brown University, Providence,
Rhode Island 02912-1843, USA}
\author{Aniket Bhattacharya}
\affiliation{Department of Physics, University of Central Florida, Orlando,
Florida 32816-2385, USA}

\date{\today}
\begin{abstract}
We investigate the influence of polymer-pore interactions on the
translocation dynamics using 2D Langevin dynamics simulations. An
attractive interaction can greatly improve translocation
probability. At the same time, it also increases translocation time
slowly for weak attraction while exponential dependence is observed
for strong attraction. For fixed driving force and chain length the
histogram of translocation time has a transition from Gaussian
distribution to long-tailed distribution with increasing attraction.
Under a weak driving force and a strong attractive force, both the
translocation time and the residence time in the pore show a
non-monotonic behavior as a function of the chain length. Our
simulations results are in good agreement with recent experimental
data.
\end{abstract}

\maketitle

The transport of a polymer through a nanopore plays a critical role in
numerous biological processes, such as DNA and RNA translocation across
nuclear pores, protein transport through membrane channels, and virus
injection. For a polymer threading through a nanopore, loss of available
configurations due to the geometric constriction leads to an effective
entropic barrier. Kasianowicz \textit{et al.}~\cite{Kasianowicz}
demonstrated that an electric field can drive single-stranded DNA and RNA
molecules through the water-filled $\alpha$-hemolysin channel and that the
passage of each molecule is signaled by a blockade in the channel current.
These observations can directly be used to characterize the polymer length.
Due to various potential technological applications~\cite{Kasianowicz,Meller03},
such as rapid DNA sequencing, gene therapy and controlled drug delivery, 
the polymer translocation has become a subject of intensive experimental
~\cite{Akeson,Meller00,Meller01,Meller02,Henrickson,Sauer,Krasilnikov,Storm}
and theoretical
~\cite{Storm,Sung,Muthukumar99,Lubensky,Metzler,Ambj3,Chuang,Kantor,Milchev,
Luo1,Luo2,Luo3,Huopaniemi1,Tian,Matysiak} studies.

As to translocation, one of the basic questions concerns the
dependence of the translocation time $\tau$ on the system parameters
such as  the polymer  chain length
$N$~\cite{Meller01,Meller02,Storm,Sung,Muthukumar99,
Lubensky,Chuang,Kantor,Milchev,Luo1,Luo2,Luo3,Huopaniemi1,Tian,Matysiak},
sequence and secondary
structure~\cite{Akeson,Meller00,Meller02,Luo3}, pore length $L$ and
pore width $W$~\cite{Luo1}, driving force
$F$~\cite{Meller01,Meller02,Henrickson,Sauer,Kantor,Luo2,Huopaniemi1,Tian,Matysiak},
and polymer-pore
interaction~\cite{Meller00,Meller02,Krasilnikov,Lubensky,Tian}.

In a recent experiment,~\cite{Meller00,Meller02} striking
differences were found for the translocation time distribution of
polydeoxyadenylic acid (poly(dA)$_{100}$) and polydeoxycytidylic
acid (poly(dC)$_{100}$) DNA molecules. The origin of the different
behavior was  attributed to stronger attractive interaction of
poly(dA) with the pore. Also, recently Krasilnikov \textit{et
al}.~\cite{Krasilnikov} have investigated the dynamics of single
poly (ethylene glycol) (PEG) molecules in the $\alpha$-hemolysin
channel in the limit of a strong attractive polymer-pore attraction.
The result for the residence time in the channel shows a novel
non-monotonic behavior as a function of the molecular weight.

On the theoretical front, not only the quantitative but also the
qualitative picture of the polymer-nanopore interactions is still
elusive. Based on a Smoluchowski equation with a phenomenological
microscopic potential to describe the polymer-pore interactions,
Lubensky and Nelson~\cite{Lubensky} captured the main ingredients of
the translocation process. However, when comparing with experiments,
their model is not sufficient. Numerically, Tian and
Smith~\cite{Tian} found that attraction facilitates the
translocation process by shortening the translocation time, which
contradicts experimental findings~\cite{Meller00,Meller02}.

To this end, in this letter we use Langevin dynamics (LD) to
investigate the influence of polymer-pore interactions on
translocation. In our numerical simulations, the polymer chains are
modeled as bead-spring chains of Lennard-Jones (LJ) particles with
the Finite Extension Nonlinear Elastic (FENE) potential. Excluded
volume interaction between monomers is modeled by a short range
repulsive LJ potential: $U_{LJ} (r)=4\epsilon
[{(\frac{\sigma}{r})}^{12}-{(\frac{\sigma} {r})}^6]+\epsilon$ for
$r\le 2^{1/6}\sigma$ and 0 for $r>2^{1/6}\sigma$. Here, $\sigma$ is
the diameter of a monomer, and $\epsilon$ is the depth of the
potential. The connectivity between neighboring monomers is modeled
as a FENE spring with $U_{FENE}
(r)=-\frac{1}{2}kR_0^2\ln(1-r^2/R_0^2)$, where $r$ is the distance
between consecutive monomers, $k$ is the spring constant and $R_0$
is the maximum allowed separation between connected monomers.

\begin{figure}
  \includegraphics*[width=\figurewidth]{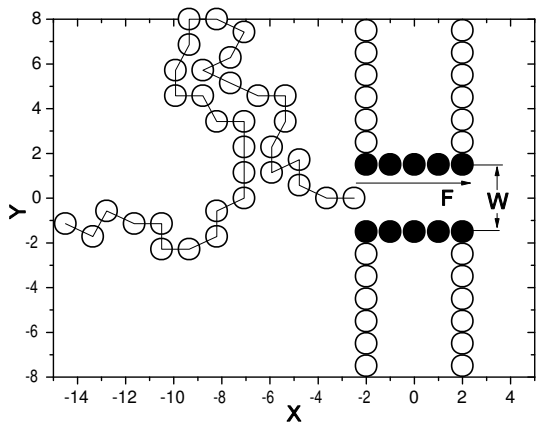}
\caption{ A schematic representation of the system. The pore length
$L=5 $ and the pore width $W=3$ ( See text for units)
        }
 \label{Fig1}
\end{figure}

We consider a 2D geometry as shown in Fig. \ref{Fig1}, where the
wall in the $y$ direction is described as stationary particles
within a distance $\sigma$ from each other. The pore of length $L$
and width $W$ in the center of the wall is composed of stationary
black particles.
Between all monomer-wall particle pairs, there exist the same short
range repulsive LJ interaction as described above.
The pore-monomer interaction is modeled by a LJ potential with
a cutoff of $2.5\sigma$ and interaction strength $\epsilon_{pm}$.
This interaction can be either attractive or repulsive depending on
the position of the monomer from the pore particles.
We have also performed numerical calculation for the case of a pure
short range repulsive LJ potential for the pore-monomer interaction.
As expected, the results for the long range LJ pore-monomer interaction
approaches that for the pure repulsive pore-monomer interaction in the
limit $\epsilon_{pm}\rightarrow 0$.
In the Langevin dynamics simulation, each monomer is subjected to
conservative, frictional, and random forces, respectively,
with~\cite{Allen} $m{\bf \ddot
{r}}_i =-{\bf \nabla}({U}_{LJ}+{U}_{FENE})+{\bf F}_{\textrm{ext}}
-\xi {\bf v}_i + {\bf F}_i^R$, where $m$ is the monomer's mass,
$\xi$ is the friction coefficient, ${\bf v}_i$ is the monomer's
velocity, and ${\bf F}_i^R$ is the random force which satisfies the
fluctuation-dissipation theorem.
The external force is expressed as ${\bf F}_{\textrm{ext}}=F\hat{x}$,
where $F$ is the external force strength exerted on the monomers in the pore,
and $\hat{x}$ is a unit vector in the direction along the pore axis.

In the present work, we use the LJ parameters $\epsilon$ and
$\sigma$ and the monomer mass $m$ to fix the energy, length and
mass scales respectively. Time scale is then given by
$t_{LJ}=(m\sigma^2/\epsilon)^{1/2}$ The dimensionless parameters in
our simulations are $R_0=2$, $k/m=7$, $k_{B}T=1.2$, and $\xi/m=0.7$.
For the pore, we set $L=5$ unless otherwise stated. A choice of
$W=3$ ensures that the polymer encounters an attractive force inside
the pore. We have checked that a choice of $W=4$ yields similar 
results. The driving force $F$ is set between $0.5$ and $2.0$, 
which correspond to the range of voltages used in the 
experiments~\cite{Kasianowicz,Meller01}. 
The Langevin equation is integrated in time by a method described by
Ermak and Buckholtz~\cite{Ermak} in 2D.
Initially, the first monomer of the chain is placed in the entrance of
the pore, while the remaining monomers are under thermal collisions described
by the Langevin thermostat to obtain an equilibrium configuration.
The translocation time is defined as
the time interval between the entrance of the first segment into the
pore and the exit of the last segment. Typically, we average our
data over 2000 independent runs.

The translocation probability, $P_{trans}$, is calculated as the
fraction of runs leading to successful translocation at given
conditions. Fig. 2(a) shows $P_{trans}$ as a function of
$\epsilon_{pm}$ for $N=128$ under different driving forces.
Specifically, the numerical results clearly show two different
regimes. With increasing $\epsilon_{pm}$, $P_{trans}$ increases
rapidly first, and then slowly approaches saturation at larger
$\epsilon_{pm}$.
It is known that attractive interaction with the
channel can facilitate the translocation of metabolite molecule
across cellular and organelle membranes~\cite{Berez}. Here, our
results show that polymer translocation through the attractive
nanopore shares the same character.

\begin{figure}
  \includegraphics*[width=\figurewidth]{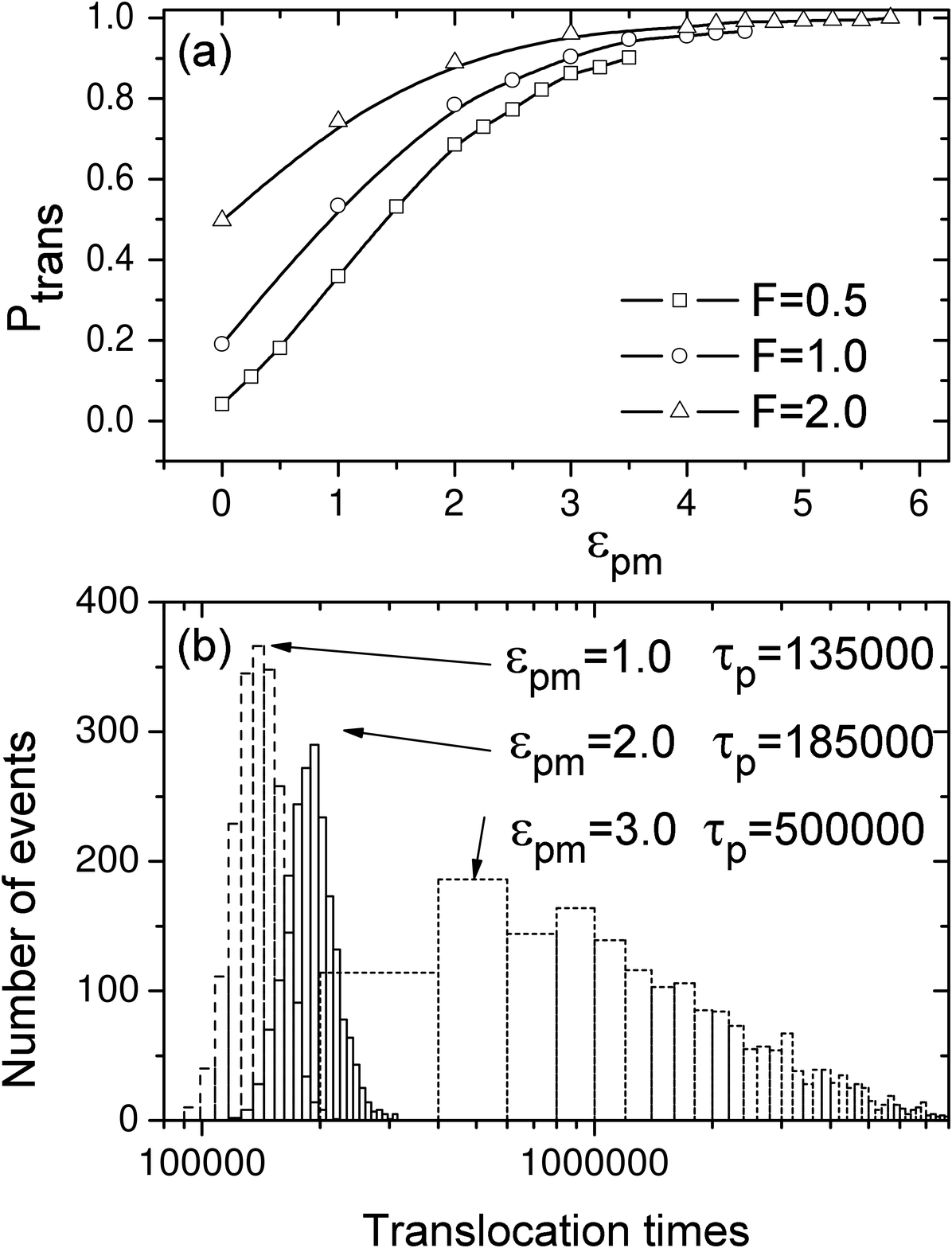}
\caption{
  (a) The translocation probability as a function of the attractive
strength for different driving forces.
  (b) The distribution of translocation time for different attractive
strengths under the driving force $F=0.5$. The chain length $N=128$.
The data point at $\epsilon_{pm} =0$ corresponds to a pure repulsive
pore-monomer interaction
        }
 \label{Fig2}
\end{figure}

Experimentally, Meller \textit{et al.}~\cite{Meller00,Meller02} have
investigated the translocation of homepolynucleotides of different
bases: poly(dA)$_{100}$ and poly(dC)$_{100}$. The translocation time
distributions in both cases are well approximated by fast-growing
Gaussian for translocation time lower than the most probable value
$\tau_p$ and falling exponentials for translocation time larger than
$\tau_p$. The decay time scale for poly(dA) is found to be much
longer than poly(dC), by a factor of $\sim 7$. There exists also a
large difference between the value of $\tau_p$ for the two,
corresponding to 1.2 $\mu s$/base for poly(dC), and 3.3 $\mu s$/base
for poly(dA). These differences have been attributed to the base
specific nucleotide-pore interactions, with the adenines having a
stronger attractive interaction with the pore as compared with
cytosines.

In our numerical results, we have found that indeed for
$\epsilon_{pm}$=2 and 3, the attractive potential has a marked
impact on the shape of the histogram of the translocation time as
shown in Fig. 2(b). The shape of the histogram changes from a nearly
Gaussian below the most probable value to a long exponential tail.
The value of $\tau_p$ as well as the characteristic decaying
time scale increases with $\epsilon_{pm}$.
These findings are in excellent agreement with the experimental
observation of Meller \textit{et al.}~\cite{Meller00,Meller02}, and
provide further support that the base specific interaction with the
pore plays a pivotal role in the translocation dynamics of single-stranded
DNA and RNA molecules.

\begin{figure}
  \includegraphics*[width=\figurewidth]{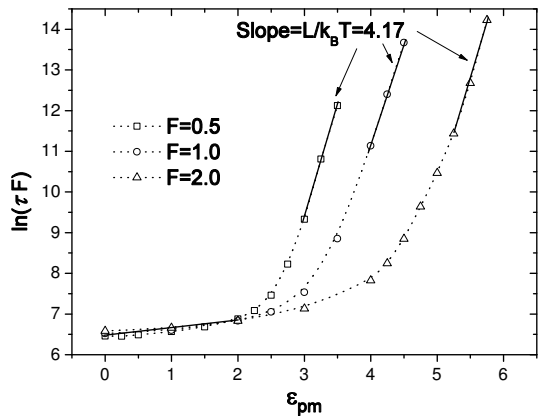}
\caption{ Translocation time as a function of the attractive
strength for different driving forces. The chain length $N=128$.
Here, $\epsilon_{pm} =0$ corresponds to a pure repulsive pore-monomer
interaction}
\label{Fig3}
\end{figure}

Fig. 3 shows the calculated $\tau F$ as a function of
$\epsilon_{pm}$ for $N=128$ under different driving forces.
Initially, $\tau$ increases very slowly with increasing
$\epsilon_{pm}$. Then, it   crosses over to a different regime and
increases sharply to the asymptotic behavior $\tau F \sim
e^{L/k_BT}$. The crossover threshold value of $\epsilon_{pm}$
increases with increasing $F$. Surprisingly, previous numerical
work~\cite{Tian} failed to capture the essential feature that the
translocation time increases with increasing attractive base-pore
interaction.

An important element of our analysis is the fact that the
translocation time can be written as $\tau \sim \tau_1 + \tau_2 +
\tau_3$, where $\tau_1$, $\tau_2$ and $\tau_3$ correspond to initial
filling of the pore, transfer of the DNA from the \textit{cis} side
to the \textit{trans} side, and finally the emptying of the pore,
respectively. In the presence of the attractive pore-monomer
interaction and driving force across the pore, $\tau_1 <<
\tau_2,\;\tau_3$, while $\tau_2$ increases monotonically with $N$.
For strong attraction and intermediate values of $N$, $\tau$ is
determined mainly by $\tau_3$ related to the emptying of the pore.
This process involves a free energy difference of $\Delta
\widetilde{F}=L(\epsilon_{pm}-F\sigma/2-f(N))$ between the final and the
initial state. The term $f(N)$ here accounts for the entropic
driving force which should kick in at larger values of $N$ and
eventually saturate for very long polymers. For the region of weak
attraction below the threshold, $\Delta \widetilde{F}<0$ and the
translocation time depends weakly on $\epsilon_{pm}$. Above the
threshold when $\Delta \widetilde{F}>0$, the process is activated
with a barrier $\sim \Delta \widetilde{F}$ and increases rapidly
with increasing strength of attraction $\epsilon_{pm}$. This
accounts for the observed crossover behavior of $\tau$ as a function
of $\epsilon_{pm}$. In the weak attraction non-activated region, the
overall $\tau$ is determined mainly by $\tau_2$ and its dependence
on the driving force scales as $F^{-1}$ which comes from the
velocity dependence on $F$. However, once one enters the activated
region, the force $F$ also influences the activation barrier besides
affecting the prefactor and $\tau$ drops off with increasing $F$
much faster than the simple $F^{-1}$ behavior.

For $F=0.5$ and a pure repulsive pore-monomer interaction, we have
shown in our earlier work~\cite{Luo2,Huopaniemi1} that $\tau\sim
N^{2\nu}$ for relatively short chains and crosses over to $\tau\sim
N^{1+\nu}$ for longer chains as shown in Fig. 4,
where the Flory exponent $\nu=0.75$ in 2D~\cite{de Gennes},
and the crossover length $N_c \sim 200$. The scaling behavior for
attractive interaction strength $\epsilon_{pm}= 1$ is very similar
to the pure repulsive case. For $\epsilon_{pm}= 2$, we found that
$N_c \sim 310$. For stronger attractive strength $\epsilon_{pm}=3$,
only $\tau \sim N^{2\nu}$ is observed for the
$N$ values studied under $F=1$ and $F=2$, with no indication of
crossover behavior as shown in the insert of Fig. 4.

\begin{figure}
  \includegraphics*[width=\figurewidth]{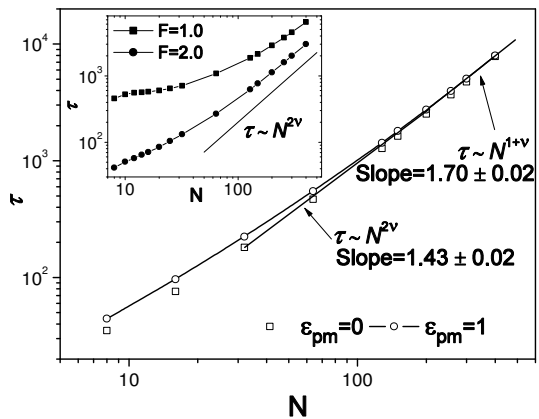}
\caption{
    Translocation time as a function of the chain length for
$\epsilon_{pm}=0$ and $\epsilon_{pm}=1$ with $F=0.5$.
The insert shows results for $\epsilon_{pm}=3$ with $F=1.0$ and $F=2.0$.
        }
 \label{Fig4}
\end{figure}

Under a strong attractive force with $\epsilon_{pm}=3$ and a weak
driving force $F=0.5$, the translocation time $\tau$ has a
qualitatively different dependence on $N$ as compared with the pure
repulsive or weak attractive pore interaction. It has a novel
non-monotonic behavior with a rapid increase to a maximum at $N \sim
14$, followed by a decrease for $14<N<32$and an increases again for
$N>32$ as shown in Fig. 5(a). This can be understood by considering
the different N dependence of $\tau_1$, $\tau_2$ and $\tau_3$ in the
strong attraction limit. For small and intermediate values of $N$,
$\tau$ is dominated by $\tau_3$. Here the entropic factor $f(N)$ in
the barrier for $\tau_3$ fights against the simple power law
increase in the prefactor accounting for the number of monomers
needed to cross the pore. This leads to an initial increase of
$\tau$ to a maximum value followed by a subsequent decrease.
Eventually, for larger $N$, the $\tau_2$ process (which approaches
$N^{2\nu}$ asymptotically) takes over, leading to the increase of
$\tau$ with increasing $N$ again.

For this case, we found that there is about $20\%$ of the total
translocation processes in which the polymer enters and reexits the
\textit{cis} side of the pore. It is useful to define an additional
residence time $\tau_r$ as the weighted sum of the translocation
time and the return time which corresponds to the experimentally
measured blockage time. For the case with no external driving
force, the translocation probability is very small and the residence
time is almost all due to return events. We have calculated the
residence time $\tau_r$ for $F=0$ and $\epsilon_{pm}=3$ and the
result is shown in Fig. 5(b). The $N$ dependence here is again
non-monotonic similar to the translocation time for $F=0.5$ except
for the absence of the eventual increase at the large $N$ limit, due
to the absence of the $\tau_2$ contribution for the return process.
Our numerical result of $\tau_r$ for $F=0$ is in good agreement with
recent experimental data of Krasilnikov \textit{et
al}.~\cite{Krasilnikov} in which the residence time of neutral PEG
molecule in $\alpha$-Hemolysin pore was measured.

\begin{figure}
\includegraphics*[width=\figurewidth]{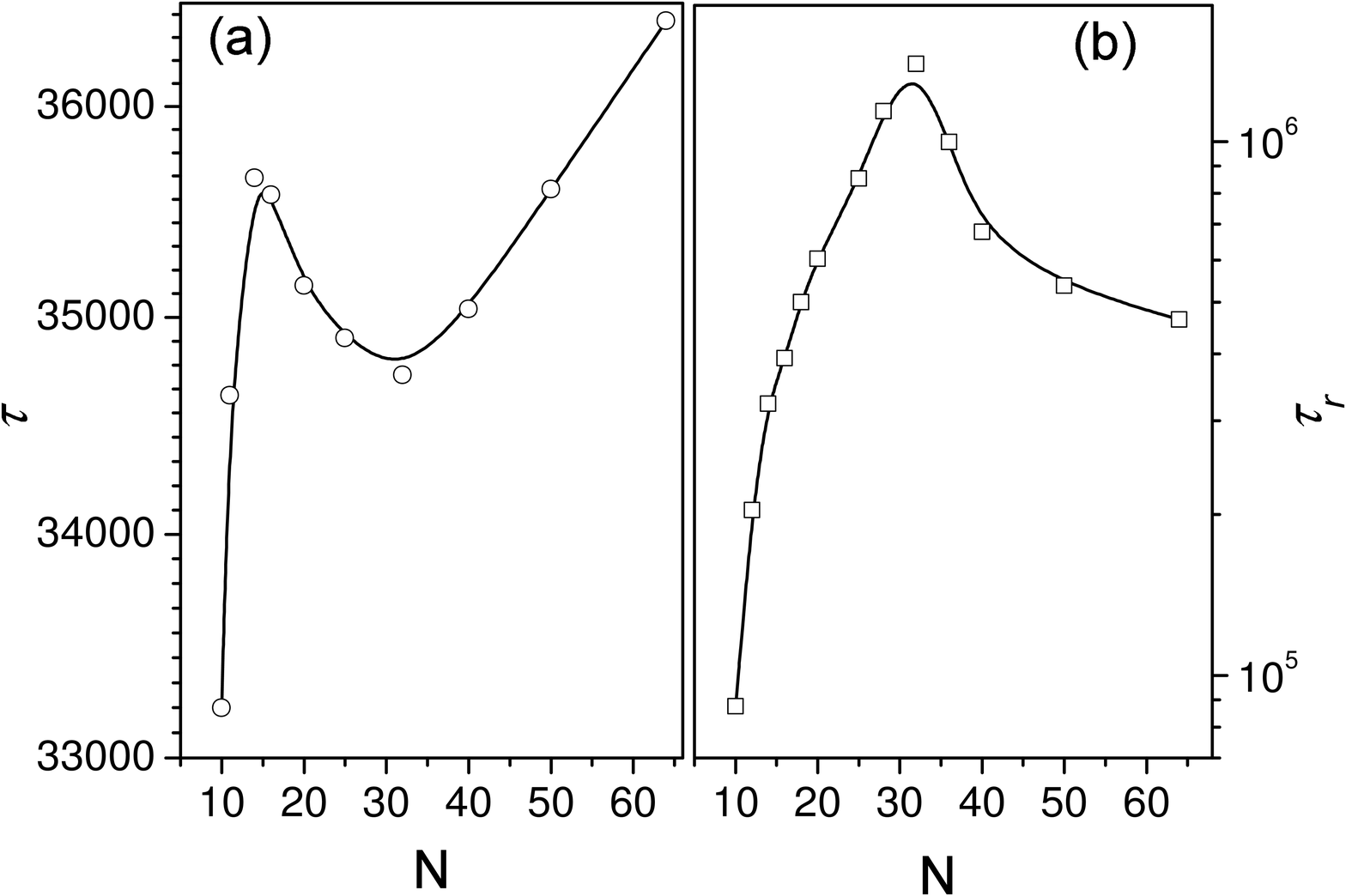}
\caption{
(a) Translocation time $\tau$ as a function of the chain length for
$\epsilon_{pm}=3$ and $F=0.5$.
(b) Residence time $\tau_r$ as a function of the chain length for
$\epsilon_{pm}=3$ and $F=0$.}
\label{Fig5}
\end{figure}

To summarize, we have investigated the influence of attractive
polymer-pore interactions on the translocation dynamics via
numerical simulation studies of a simple course grained model. Our
results are in good agreement with recent experimental data for
driven translocation of poly(dA)and poly(dC) molecules, and for the
 blockage time study of poly(ethylene glycol) molecule through $\alpha$-Hemolysin pore.
 They
clearly demonstrate the important role of polymer-pore interaction
factor in the translocation dynamics.

\begin{acknowledgments}
This work has been supported in part by The Academy of Finland
through its Center of Excellence (COMP) and TransPoly Consortium grants.
\end{acknowledgments}

\end{document}